\documentclass[12pt]{iopart}
\usepackage{iopams}  
\usepackage{graphics}
\usepackage{epsfig}
\begin{document}

\title{Monte Carlo simulations
of the L1$_0$ 
long-range order  relaxation in
dimensionally reduced systems
}

\author{Mohammed Allalen\dag \footnote[4], Tarik Mehaddene\ddag  
{To
whom correspondence should be addressed (allalen@lrz.de)}, Hamid Bouzar\S
}

\address{\dag\ Universit\"at Osnabr\"uck, Fachbereich Physik,
D-49069 Osnabr\"uck, Germany \\ \ddag\ Physik-Department E13, Technische Universit\"at M\"unchen, 85747 Garching, Germany\\ \S \  LPCQ, University Mouloud Mammeri, 15000 Tizi-Ouzou, Algeria }

\begin{abstract}
Monte Carlo simulations have been performed to investigate the relaxation of the L1$_0$ 
long-range order in
dimensionally reduced systems.
The
effect  of the number of (001)-type
 monatomic layers and of the  
pair interaction energies on these kinetics has been examined.
The vacancy
 migration energies have been deduced
from the Arrhenius plots 
of the relaxation times. 
A 
substantial increase in the migration energy 
 for small film thickness is observed.
 The results 
agree 
with previous Monte Carlo simulations 
 and with recent experimental results in L1$_0$ thin films and multilayers.    

\end{abstract}


\submitto{\JPCM}

\maketitle

\section{Introduction}
Binary ferromagnetic materials with the L1$_{0}$ structure are promising candidates for use as magneto-optical recording media 
thanks to their high 
magnetic anisotropy. The L1$_0$ structure (also called the CuAu structure) is based on the face centered 
lattice of tetragonal symmetry. It is build by 
alternating $\{001\}$ pure planes of each type of atoms. Three variants are present in the  L1$_{0}$ structure: x-, y- and z-variants. The z-variant is build when the (001) pure atomic planes are stacked along the [001] direction, whereas the x- and y- variant are obtained when the pure atomic planes are stacked along the [100] and [010] directions respectively. Among the most investigated systems are  Co-Pt, Fe-Pt,
 and Fe-Pd.
They present a tetragonal distortion of the lattice ($c/a <$ 1) \cite{9,13,41}
and  exhibits a uniaxial magnetocrystalline anisotropy in the range $K_{1}=1-10$~MJ.m$^3$ 
with
 the $c$-axis as the easy magnetisation axis.
The discovery of the interesting properties of these materials dates back to the 1930's. They have been, however, 
 the topic of many intensive experimental and theoretical works in the last decades (see \cite{veroo,zfmk} and references therein).
 A good knowledge of the ordering processes and of its dynamics is 
 a necessary step in any extensive research on those systems. 
\\
\\
A systematic study of the physical, thermodynamic and kinetic
properties in the Fe--Pd, Fe--Pt and Co--Pt ordered alloys is now in progress.
  Atomic migration has been
 investigated using experimental resistivity measurements and lattice dynamics
 to deduce the atomic migration energies
 at different temperatures and  states of atomic order. The study is almost completed
 in FePd \cite{PRB,TM2,TM3,messad} whereas, the other two systems are still under investigation.  In parallel to these 
experimental investigations,
 numerical simulations of the thermodynamical properties and structural kinetics have been performed
 using Monte Carlo methods and molecular dynamics \cite{zfmk,moh1}. 
 L1$_0$-ordering kinetics in FePt nano-layers have been
recently simulated using a Monte Carlo method based on pair interaction energies deduced from the Cluster
Expansion method \cite{rafal}. The results indicate a clear stabilisation effect of Pt-surface on the $z$-variant of the  L1$_0$ structure.
\\
\\
In this paper, we present systematic Monte Carlo simulations
of the L1$_0$ 
long-range order  relaxation in
dimensionally reduced systems.
 For this purpose the isothermal relaxation of the 
long-range order (LRO) parameter is investigated in the $z$-variant of the  L1$_0$ structure,
 with different numbers of (001)-type monatomic layers. We first present the effect of the film thickness on 
 the order-disorder  transition for different ordering energies. To get more insight into the order-order kinetics, the isothermal relaxation of a short-range order parameter, namely the anti-site pair correlation (APC) parameter is simulated along with the LRO parameter. Finally, the Arrhenius plots of the relaxation times are used to deduce the vacancy migration energy for different film thicknesses.  
The present work attempts to know to which extent are influenced the kinetics in dimensionally 
reduced L1$_0$ compounds
in comparison to processes taking place in bulk systems.

\section{Simulation Method}
Stochastic techniques, based on random processes, such as Monte Carlo methods have been used to solve
a wide variety of problems \cite{Bin:S79}. They have been established as a powerful technique to investigate the
 kinetics relaxation of the long-range order in intermetallic \cite{Yal:JMR95,oramus,scripta}. 
\\
\\
Dimensionally reduced  L1$_{0}$ systems have been modeled 
 by a set of $N_{A}$ atoms $A$  and $N_{B}$ atoms $B$ distributed on a perfect face-centered-cubic rigid lattice 
 with a linear size  $L=64$ along the [100] and [010] directions and variable thickness $M$ along the [001] direction.
 To conform to the stoichiometry of the L1$_{0}$ structure, there are as many $A$ as $B$ atoms.
 Periodic boundary conditions have been used along $x$ and $y$-axis whereas
these have been removed  along the $z$-axis. As a result of symmetry breaking, relaxation and
 reconstruction processes often take place at the free surfaces of real thin-films inducing 
local defects and building extra units or clusters. Surface-induced defects being beyond the 
scope of the present study, we restrict ourself to flat and ideal surfaces.
 Assuming interactions up to next nearest-neighbours only, the total energy of the system can 
be expressed in a general $ABv$ model by an Ising Hamiltonian \cite{gautier}
\begin{equation}
{\cal H}=\frac{1}{2}\sum_{i,j}V_{ij}\sigma_{i}\sigma_{j}
\end{equation}
where the sum extends over all the nearest and the next nearest-neighbour pairs and $i$ and $j$ are 
generic indexes sweeping all the lattice. The occupation operator on site $i$, $ \sigma_{i}$ can take
 the values: +1, -1, and 0 if site $i$ is occupied by an $A$ atom, a $B$ atom, or a vacancy $v$, 
respectively. $V_{ij}$ are the  pair interaction energies between atoms at sites $i$ and $j$.
 For convenience the pair interactions can be labeled by the shell index $n$, which in the present work 
can take the value 1 and 2 for the nearest-neighbours and the next nearest-neighbours, respectively.
 Following earlier works in bulk L1$_0$ compounds \cite{Kerrache}, 
phenomenological pair interaction energies are
 used with a constant value of the first pair interaction $V_{1}=20meV$ and different 
values of the second pair interaction V$_{2}$. We have chosen to restrict 
our study to symmetric interaction energies and neglect the vacancy-atom interactions:
\begin{equation}
V_{vA}=V_{vB}=0 \ ; \ \ V_{AA}=V_{BB}=-V_{AB}
\end{equation} 
Vacancy-atom interactions are usually considered to take into account local distortions and relaxation around vacancies. However, the static lattice displacement has been measured by neutron diffuse scattering in an L1$_{0}$-FePd by one of the present authors 
and has been found to be 
weak \cite{TM2}. The use of symmetric interactions greatly reduces the number of simulation parameters without 
changing the order-disorder transition temperature \cite{Yal:JMR95}. The use of such "toy model" has been widely and successfully used to investigate the kinetics relaxation in binary systems \cite{zfmk,Yal:JMR95,oramus}. The model is based on the atomic mechanism of order relaxation in dense phases \cite{petry}:
  vacancy-atom exchange between nearest-neighbour sites.  The simulation starts with a perfect
 L1$_0$ ordered crystal in which one of the two sub-lattices
(sub-lattice $\alpha$) is occupied by A atoms and the other (sub-lattice $\beta$) by B atoms.
 To keep close to real systems and avoid any interaction effect between vacancies, only
 a single vacancy is introduced at random in the simulation box. 
 The elementary Monte Carlo step is the following: one of
the vacancy neighbours (A or B atom) is randomly chosen, the energy balance $\Delta {\cal H}$ of
the atom-vacancy exchange  before and after the jump is evaluated with the  Ising Hamiltonian (Eq. 1), 
considering the two nearest-neighbour shells of the initial and final atomic positions.
The jump  is performed if the Glauber probability \cite{Glauber}
\begin{equation}
P(\Delta {\cal H})=\frac{1}{[1+exp(\Delta {\cal H}/k_{B}T)]} 
\end{equation}
is larger than a
random number between $0$ and $1$. 
This corresponds to averaging the result over a large number of reversal jump attempts, the
sum of the probabilities of the jump and its  reversal being equal to $1$.
\\
\\ 
The configuration of the system was analysed, at regular time intervals, 
by calculating two  parameters: (i) an effective LRO parameter $\eta=2(2N_{\rm A}^{\alpha}-N_{\rm A})/(N_{\rm sites}-1)$, 
where $N_{\rm A}^{\alpha}$ is the number of A atoms on the $\alpha$ sub-lattice and
$N_{\rm A}$ the total number of A atoms. (ii)
 $APC=\frac{N_{AB}^{\alpha\beta}}{N_{B}^{\alpha}}$,
 where $N_{AB}^{\alpha\beta}$ is the number of nearest-neighbour pairs of antisites and $N_{B}^{\alpha}$
id the number of $\alpha$-antisite. The time scale being the number of jump attempts. One should keep in mind that the Monte Carlo time we refer to here is a "raw" time. Its mapping to  physical time scale in seconds, in a system-specific study, requires the knowledge of the physical properties e.g. the vacancy concentration at different temperatures, diffusion coefficient etc. 
 The relaxation of the LRO  and the APC parameters towards their equilibrium values has been 
simulated in the $z$-variant, with (001)-type monoatomic layers of the L1$_{0}$ structure
 for different film thicknesses and different values of
 $k=V_{2}/V_{1}$. For each temperature, the evolution of  $\eta$  is followed until the system reaches 
equilibrium. We have chosen to stop when the simulation time was at least 5 times greater than  the longer relaxation time of the system.  The number of (001) layers has been varied from 8 up to
 190 leading to  more than 2000 simulations. Due to size effects, simulations of small film thicknesses
 suffered from poor statistics and very long relaxation times. All calculations have been performed on Pentium 4 processors using a Fortran 77 program.

\section{Order-disorder transition}

 The equilibrium value of the LRO parameter  $\eta_{eq}$ {\it versus} temperature is plotted
 in Figure 1 for different film thicknesses in the case  $k=-0.2$. As expected, the order-disorder
 transition temperature, T$_{c}$, decreases with decreasing  film-thicknesses as a result of the
 decrease in the  
 relative number 
 of atomic bounds caused by the symmetry breaking at  the free surfaces.
 This behaviour is well seen in Figure 2 in which the reduced critical temperature
 $T_{c}/T_{c}^{bulk}$ is reported against the film-thickness $M$ for different values of
 $k$. $T_{c}^{bulk}$ being defined as the transition temperature for film-thicknesses greater
 than 128. The reduced transition temperature falls on a unique curve for all the  pair
 interaction values, showing a plateau for $M\ge 64$, reflecting the bulk behaviour of the
 system. 
The extrapolation of the curve towards very thin films gives,
 in the limit $M=1$,  zero. Indeed, in this limit,
the system  reduces to one monoatomic plane containing only one 
kind of atoms. 
The system becomes then unstable
regardless of  the sign of  first and second 
nearest-neighbours  pair interactions. 
\\
\\
For small film-thicknesses the  order-disorder, 
 transition is found to be of first order type  and  the transition is 
clearly discontinuous. The calculated transition temperatures for $M\ge 64$ for different values
 of  $k$ agree quite well with previous Monte Carlo simulation of the LRO relaxation in L1$_{0}$
 bulk compounds \cite{kerrache}.
  For low and high film thicknesses as well, the transition 
temperature has been found, as expected, correlated to the ratio $k=V_{2}/V_{1}$. 
\\
\\
Figure 3 shows a typical snapshot of two neighbouring (010) atomic planes obtained in the equilibrium
 state of a film with 32 atomic layers at $T=550~K$. It is clearly seen that the disordering 
takes place first at the free surfaces and  then grows into the film. Nucleation of the $x$ and the
 $y$-variants of the L1$_{0}$ structure are also seen.

\section{Two time scales "order-order" relaxations}
In good agreement with the predictions of the path probability method \cite{ppm} and with
many experimental and computational results in intermetallics \cite{zfmk,oramus,scripta,31,32,33,34}, 
 the relaxations of the LRO have been found to be  well fitted with  the sum of two exponentials
 yielding a long $\tau_{l}$ and short $\tau_{s}$ relaxation time:
\begin{equation}
\frac{\eta (t)-\eta_{eq}}{\eta (t=0)-\eta_{eq}}=C \ exp\Big(-\frac{t}{\tau_{s}}\Big)+(1-C) \ exp\Big(-\frac{t}{\tau_{l}}\Big)
\end{equation} 
 $0\le C\le 1$. Figure 4 shows an example of the LRO relaxation towards the
 equilibrium value $\eta_{eq}=0.72$  obtained for $M=32$ with $k=-0.2$ and
 its fit using  two exponentials.  The  coefficient values $C$ of the
 fast relaxation process deduced from   fits using Eq. 4 for
 different film thicknesses and different  values of $k$  are depicted 
 in Fig.~5. Note that presenting the results as a  function 
of the LRO is equivalent to presenting them as a function 
of temperature. In good agreement with previous Monte Carlo simulation of the
 kinetics relaxation in bulk L1$_{0}$-FePd \cite{scripta}, the fast process 
is dominant for $M=64$. Its contribution to the overall
 relaxation however decreases from $C \ge 0.9$ down to $0.78$ for $M=32$
 ($k=-0.2$, $\eta_{eq}=0.90$). For low enough thicknesses and high ordering
 states, the fast process  smears out and the slow one becomes 
dominant ($C \le 0.4$).  A detailed study in the L1$_{2}$ phase has shown
 that the short relaxation time 
is related to the formation of the nearest-neighbour antisite pairs, whereas the 
longer one is related to the uncoupling of these antisite pairs \cite{oramus}.
The L1$_{2}$ phase of AB$_{3}$ compounds is also an fcc-based structure with however
a stoichiometry proportion $N_{A}/N_{B}=1/3$. The A atoms occupy the  cubic 
cell corners and the B atoms occupy the center of faces. 
 The striking feature of Fig.~5 is that for low film thicknesses ($M=16$) the
 same final degree of order ($\eta_{eq}=0.90$) is reached through different 
processes, depending on the pair interaction energies. For $k=-0.2$ the fast
 process contributes with a fraction of $35\%$ to the overall 
relaxation whereas when $k=-0.5$ the fast process contributes more than $68\%$
 stressing definitely both geometrical and energetic effects on the feature
 of "order-order" relaxations.  To further investigate  this point, we have 
calculated the relaxation of the APC parameter towards its equilibrium value 
for different film thicknesses with nearly the same  final state of order 
($\eta_{eq}=0.90$ for $k=-0.2$ and $\eta_{eq}=0.96$ for $k=-0.5$). 
The results are depicted in Fig.~6. For low film thicknesses, namely
 $M=16$, where the contribution of the fast process is less than $40\%$,
 the early disordering stages are accompanied by a very fast increase
 of APC which reaches a maximum value before starting to decrease, while
 the effective LRO parameter decreases further. This behaviour is not  
seen in cases where the fast process is dominant, that is for film 
thicknesses $M=32,64$. The fast decrease of the LRO in the early stage
 of relaxation is due to the formation of nearest-neighbour pair of antisites
 (increase of APC in Fig.~6 ). This process is very efficient in  
quickly decreasing  the LRO. It saturates however after a while, and  further
 decrease in the LRO (slow process) is due to the uncoupling of these 
nearest-neighbour antisite pairs  which give rise to single antisite 
diffusion jumps (decrease of APC in Fig.~6 ). Both in L1$_{2}$ and 
L1$_{0}$ structures antisite atoms may easily migrate within one sublattice
 without inducing any disorder. This process was shown to be responsible of
 the slow relaxation process in   L1$_{2}$-ordered compounds due to the
 uncoupling of the nearest-neighbour antisite pairs. Such events occur in 
the L1$_{0}$ structure within the basal planes and  only four jump
 possibilities are offered, in comparison to eight in the  L1$_{2}$ structure.
 As a result, such antisite diffusion jumps are definitely rare in L1$_{0}$ 
compounds in comparison to  L1$_{2}$ compounds. This explains  the very
 low contribution of the slow process in bulk-L1$_{0}$ compounds. 
However, by reducing  the $z$-axis dimension, the fraction of 
such jumps become equal or even higher than that all possible
 jumps giving rise to a dominant slow relaxation process,  obtained in the case of
 $M=16$ (Fig.~6). 

\section{Vacancy migration energy}

The longer relaxation time, $\tau_{l}$ was found to follow an Arrhenius law yielding  
a vacancy migration energy, $E_{M}$, which has been deduced from the linear regression of the Arrhenius plots
for different  film thicknesses $M$ in the case $k=-0.2$ 
 (Figure 7). The increase in $E_{M}$ for low film thicknesses can be explained by the  fact that
the thinner the film  the smaller are the degrees of freedom of the atomic motion. The film thickness effect  is strong not only on the migration energy but also on the diffusion process itself. Depending on the film thickness, the final equilibrium state is reached by different routes  (See Fig.~6) along which  the vacancy encounters different atomic surroundings, leading to different migration energies.
 If one extrapolates  $E_{M}$ towards very low film thicknesses, namely $M=1$, we get a migration
 energy of $0.40\pm 0.05~eV$. 
Kerrache {\it et al} \cite{Kerrache}  have obtained a linear
 variation of  $E_{M}$ with $k$ for binary two-dimensional lattice.
 The linear extrapolation of their data to $k=-0.2$ leads to $E_{M}=0.37~eV$, in good agreement
 with the present deduction.  One should, however, keep in mind that, $E_{M}=0.40~eV$ is the extrapolation
 from $M=16$ towards very small thicknesses, thus the comparison is rather qualitative.
 Smaller film thicknesses are necessary for any further discussion, they were, however,
 extremely difficult to simulate within reasonable CPU time. 
\\
\\
Different experimental methods have been used to determine  activation
 energies in L1$_{0}$ FePt and CoPt thin films and multilayers. Grazing Incidence Synchrotron 
Reflection (GIRNS) \cite{ddf} has been used to measure   diffusion coefficient at different 
temperatures in FePt multilayers. Due to  the geometry of the  set-up used,  only the  
 activation energy  along the c-axis could be measured. It has been  found to be equal to $1.65~eV$,
 which differs from   extrapolation of high temperature tracer data \cite{kushida}. Resistivity
 measurement showed that the activation energy drops from $2.7~eV$ for $T\ge 830~K$ to $1.5~eV$
 for $T\le800~K$ \cite{ddf}. The latter value agrees quite well with those deduced from the
 shift with temperature of the differential scanning calorimetry peak of the order-disorder
 transition with heating rate in FePt thin films ($1.6~eV$) \cite{barmak}. The same analysis
 in CoPt thin films yields an activation energy of $2.8~eV$.
 It is believed  that the diversity of activation energies in FePt multilayers  results from the 
variety of processes involved in the relaxations. Assuming that the activation energy,
 $E_A$, is the sum of $E_{M}$ and the vacancy formation energy $E_{F}$, the simulated 
 values of $E_M$ agree qualitatively  well with the experimental data. In our
 study, we keep a constant number of vacancies, therefore we do not have access to $E_{F}$.
 The vacancy formation energy $E_{F}$ has been measured in pure Co, Pt \cite{E_F} and 
Fe \cite{lb}, it has been found equal to $1.38~eV$, $1.2~eV$ and $1.8~eV$  for Co, Pt and Fe respectively.
 Considering the simulated  $E_{M}=0.40\pm 0.05~eV$ obtained for the lowest film thickness
 $M=16$ for $k=-0.2$ in Figure 7, added to an average value of E$_{F}$ in the pure Co and Pt
 component, we get an activation energy in CoPt of $1.7~eV$,  which has to be compared to 
 $2.8~eV$, measured in equiatomic CoPt thin films. A better agreement is obtained in the case of FePt.
 With the same approach, we get E$_{A}=1.9~eV$ which agrees  well with the
 measured  activation energy in FePt thin films given above. The better agreement observed for FePt 
in comparison to CoPt is not fortuitous. Paudyal {\it et al.} have calculated the pair interaction energies in 
L1$_{0}$-CoPt and FePt using first-principle calculations \cite{pau}. They give, $V_{1}^{FePt}=19.72~meV$, $V_{2}^{FePt}=-0.51~meV$ for FePt and $V_{1}^{CoPt}=27.33~meV$, $V_{2}^{CoPt}=-0.034~meV$ for CoPt. The value  of the first pair interaction of FePt   used in our calculations is very close the calculated value.

\section{ Conclusion} 
\label{sec-5}

A Monte Carlo method was applied to investigate the relaxation of the LRO in dimensionally 
reduced  L1$_0$ compounds. The order-disorder transition has been found to be  
first-order for low film thicknesses and, as expected, its temperature strongly dependent on the pair interaction
 energies. A substantial increase in the migration energy has been obtained for low film
 thicknesses. By analysing  the relaxation of the    LRO  and the antisite pair 
correlation  parameters, we could stress that both geometrical and energetic effects play an important 
role  in the interplay between the fast and the slow relaxation processes during "order-order" relaxations.
 Despite of the simplicity of our model, which does not take into account substrate  effect and
 assumes temperature independent interaction energies, we could reproduce qualitatively
 migration energies measured in L1$_{0}$ thin film and multilayers. 

\vspace*{2cm}

{\small{
    The authors would like to thank Dr.~V.~Pierron-Bohnes ( CNRS/IPCMS Strasbourg, France) for carefully reading the first part of this manuscript and Prof.~Dr.~J\"urgen Schnack (Universit\"at Osnabr\"uck, Germany) for his corrections. .
}
}

\newpage
\vspace*{2cm}
\begin{figure}[h!]
\begin{center}
\resizebox{0.7\columnwidth}{!}{%
  \includegraphics{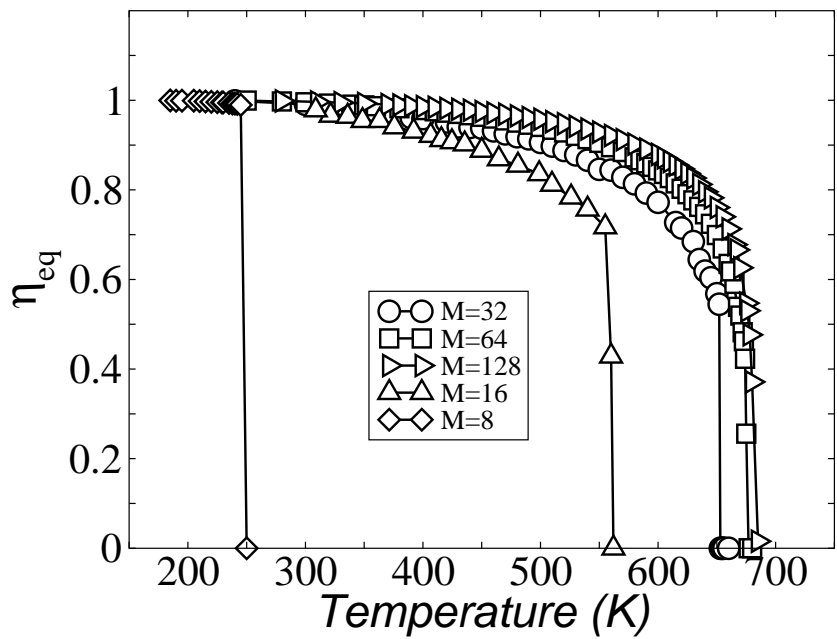}}
\end{center}
\caption{Equilibrium LRO {\it versus} temperature for different film thicknesses obtained for $k=-0.2$.}
\end{figure}

\newpage
\vspace*{2cm}
\begin{figure}[h!]
\begin{center}
\resizebox{0.8\columnwidth}{!}{%
  \includegraphics{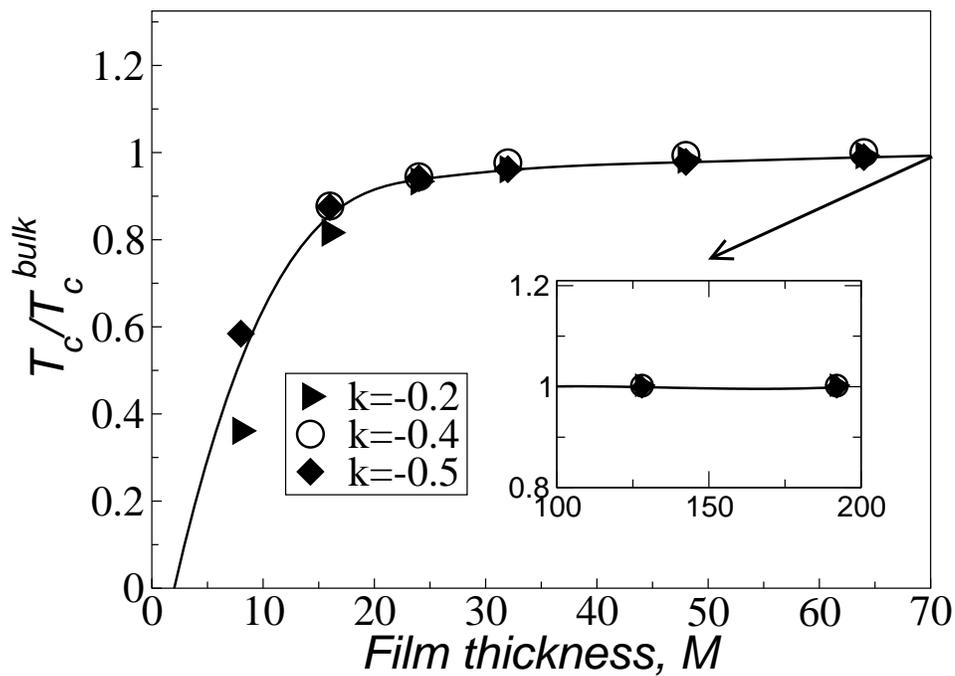}}
\end{center}
\caption{Reduced temperature {\it versus} film thickness for different values of $k$. T$_{c}^{bulk}$ being the order-disorder transition temperature for $M\ge 128$. The line is a guide for eyes. The inset shows the asymptotic behaviour for higher film thicknesses. }
\end{figure}

\newpage
\vspace*{2cm}
\begin{figure}[h!]
\begin{center}
\resizebox{0.85\columnwidth}{!}{%
  \includegraphics{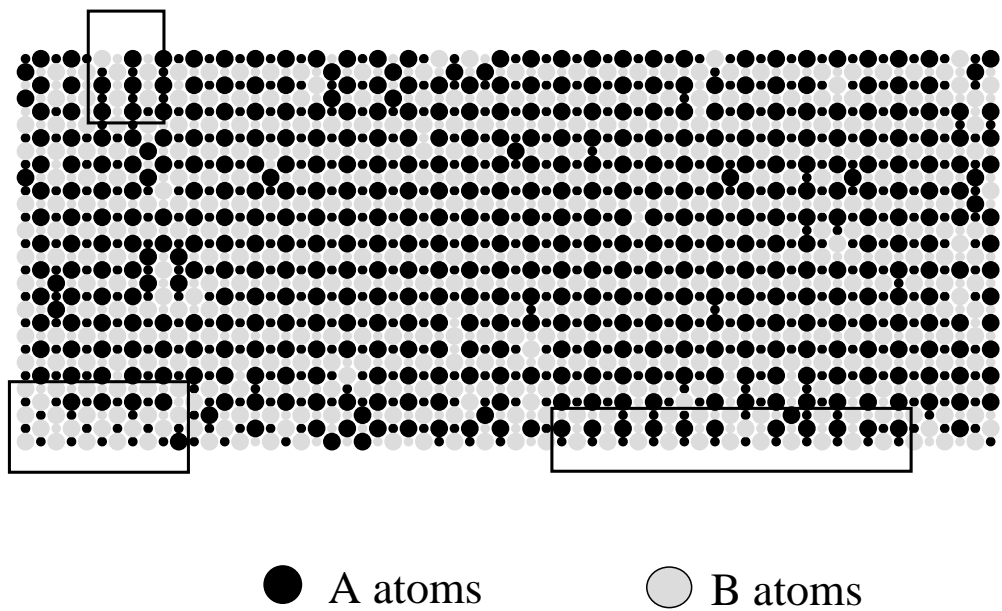}}
\end{center}
\caption{Snapshot of an (010) atomic plane (large circles) and its neighbouring plane (small circles) in an equilibrium state of  an  L1$_{0}$ thin film with 32 atomic layers at $T=550~K$ for $k=-0.2$. The boxes indicate nucleation of the $x$ (upper and down left boxes) and $y$ (lower right box) variants of the L1$_{0}$ structure.}
\end{figure}

\newpage
\vspace*{2cm}
\begin{figure}[h!]
\begin{center}
\resizebox{0.8\columnwidth}{!}{%
  \includegraphics{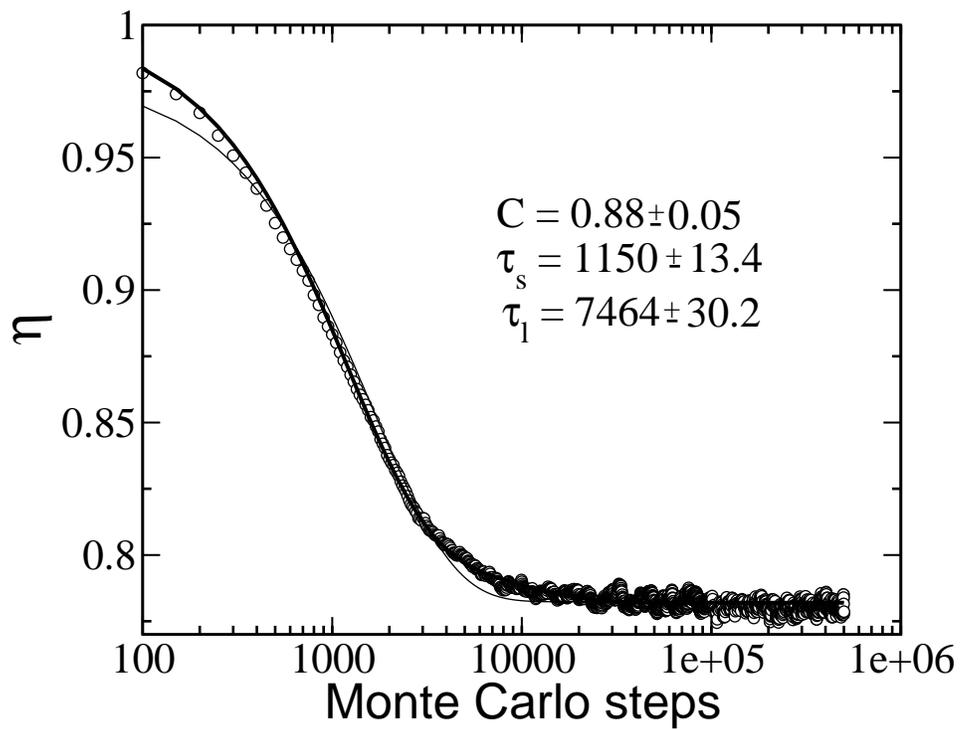}}
\end{center}
\caption{Semilog plot of the isothermal relaxation of $\eta$ obtained for $T=950~K$ and $k=-0.5$ for $M=48$ (circles) and its simulation using the sum two exponentials (thick line). For comparison, a fit using a single exponential is also 
shown (thin line).}
\end{figure}

\newpage

\begin{figure}[h!]
\vspace*{2cm}
\begin{center}
\resizebox{0.75\columnwidth}{!}{%
  \includegraphics{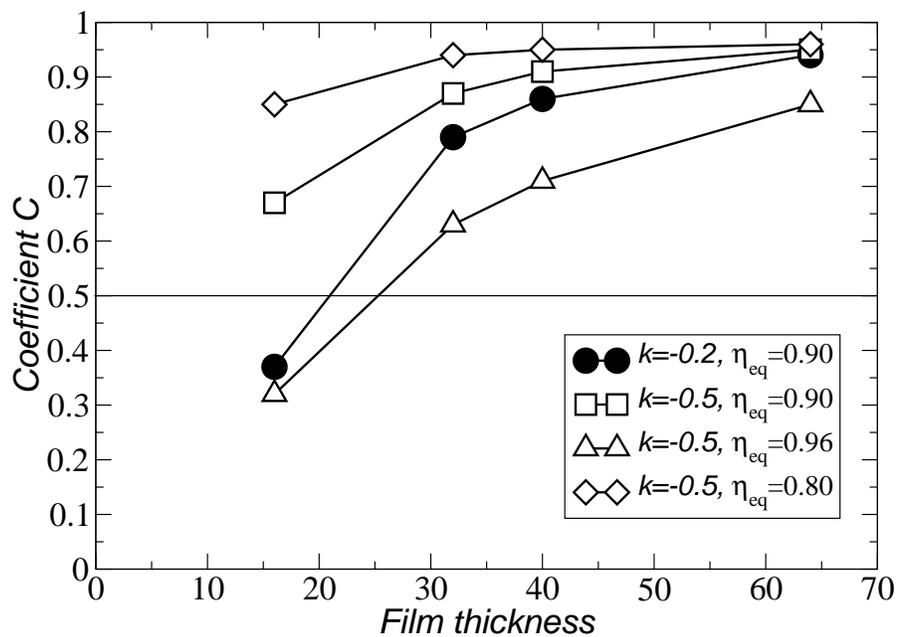}}
\end{center}
\caption{Variation of the coefficient $C$ of the fast relaxation process {\it versus} film thickness for different values of $k$ and different states of order.  Note that presenting the results as a  function 
of the LRO is equivalent to presenting the results as a function 
of temperature.}
\end{figure}

\newpage
\vspace*{2cm}
\begin{figure}[h!]
\begin{center}
\resizebox{1.1\columnwidth}{!}{%
    \includegraphics{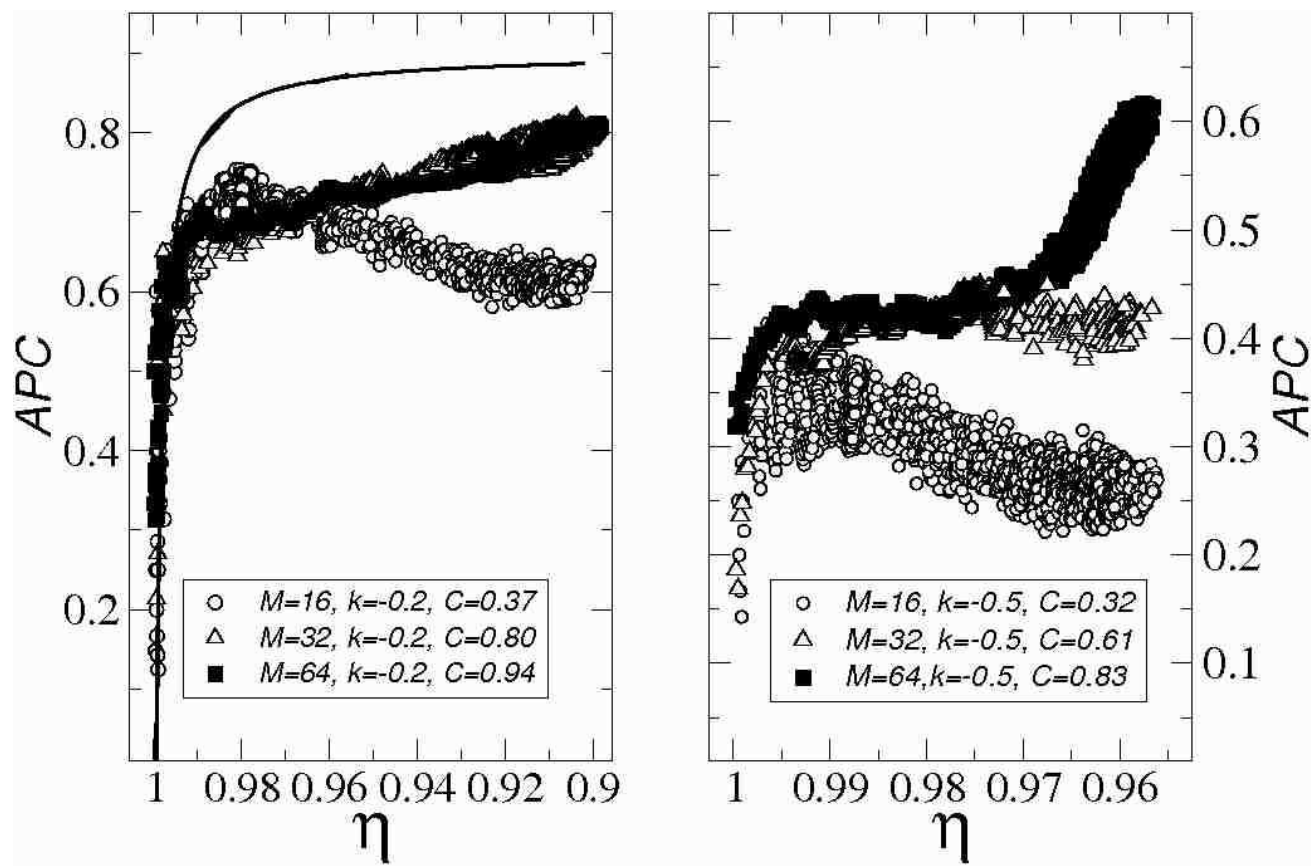}}
\end{center}
\caption{APC {\it versus} $\eta$ during the relaxation of the systems with $M=16$, $32$ and $64$ towards the equilibrium value $\eta_{eq}=0.90$ for $k=-0.2$ (left) and $\eta_{eq}=0.96$ for $k=-0.5$ (right). }
\end{figure}

\newpage
\vspace*{2cm}
\begin{figure}[h!]
\begin{center}
\resizebox{0.9\columnwidth}{!}{%
    \includegraphics{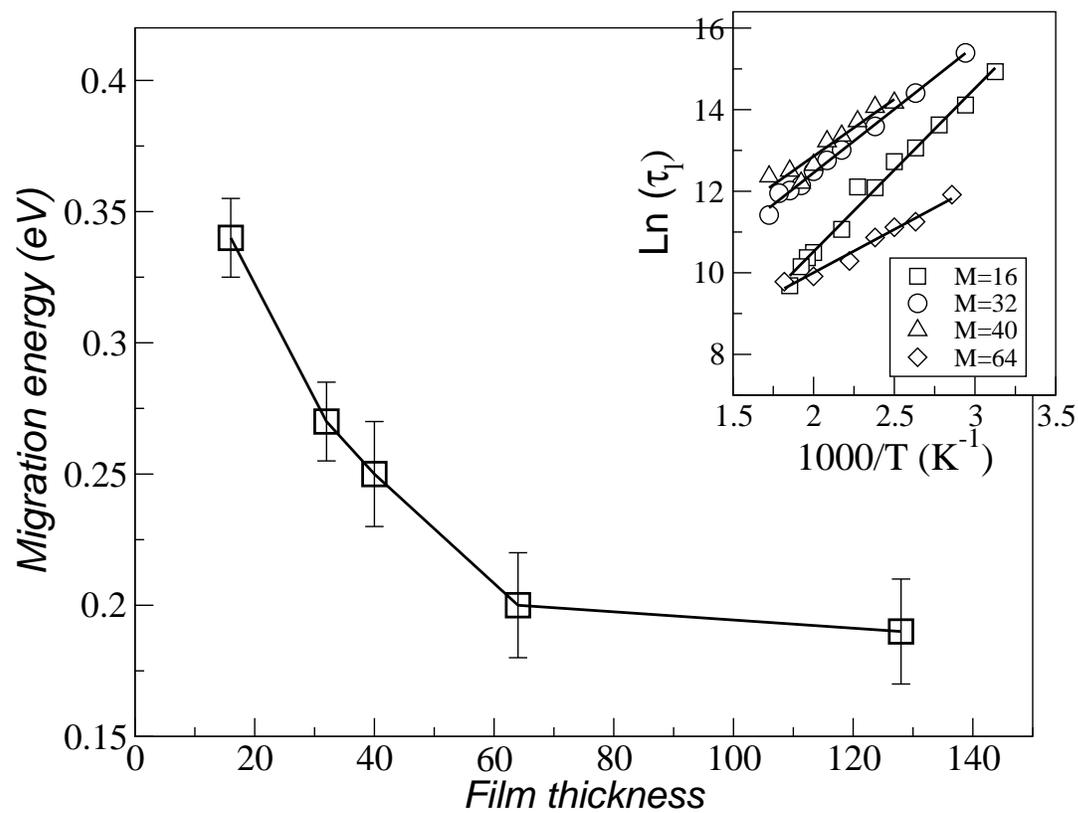}}
\end{center}
\caption{Variation of $E_M$ with the film thickness for $k=-0.2$. In inset:  Arrhenius plots of the longer relaxation times.}
\end{figure}

\end{document}